\def\ga{\mathrel{\mathpalette\fun >}}
\def\fun#1#2{\lower3.6pt\vbox{\baselineskip0pt\lineskip.9pt
 \ialign{$\mathsurround=0pt#1\hfil##\hfil$\crcr#2\crcr\sim\crcr}}}
\relax
\documentstyle[12pt]{article}
\advance\textheight by \topskip
\begin{document}
\thispagestyle{empty}
\begin{flushright}
{SU--ITP--94--2}\\
astro-ph/9402031\\
February 11, 1994
\end{flushright}
 \vskip 1.8cm
\begin{center}
{\Large\bf  MONOPOLES \\
\vskip .9cm
AS BIG AS A UNIVERSE}\\
\vskip 2.5cm
  {\bf Andrei Linde}\footnote{On leave  from:  Lebedev
Physical Institute, Moscow, Russia.\
E-mail: linde@physics.stanford.edu},
\vskip 0.05cm
Department of Physics, Stanford University, Stanford, CA 94305-4060,
USA
\end{center}
\vskip 1.5cm

{\centerline{\large ABSTRACT}}
\begin{quotation}
\vskip -0.4cm
\ \ We show that, contrary to the standard belief,  primordial
monopoles expand
exponentially during inflation in the new inflationary universe
scenario.
Moreover, inflation of monopoles continues without  an end even when
inflation
ends in the surrounding space. Therefore primordial monopoles (as
well as other
topological defects produced during inflation) can  serve as seeds
for
the process of eternal self-reproduction of inflationary universe.
 \end{quotation}

 \vfill
\newpage

\section{\label{1} Introduction}

1. One of the main purposes of inflationary cosmology was to solve
the
primordial monopole problem.  The solution was very simple: inflation
exponentially increases the distance between monopoles and makes
their density
negligibly small.
It was assumed that inflation neither can  change internal properties
of
monopoles nor can it lead to their copious production. For example,
in the
first version of the new inflation scenario based on the $SU(5)$
Coleman-Weinberg theory \cite{b16} the Hubble constant during
inflation was of
the order $10^{10}$ GeV, which is five orders of magnitude smaller
than the
mass of the $X$-boson $M_X \sim 10^{15}$ GeV. The size of a monopole
given by
$\sim M_X^{-1}$   is five orders of magnitude smaller than the
curvature of
the universe given by the size of the horizon $H^{-1}$. It seemed
obvious that
such monopoles simply could not know that the universe is curved.

This conclusion finds an independent confirmation in the calculation
of the
probability of spontaneous creation of monopoles during inflation.
According to
\cite{GuthVil}, this probability is suppressed by   a factor of $\exp
(-2\pi
m/H)$, where $m$ is the monopole mass. In the model discussed above
this
factor is given by $\sim 10^{-10^6}$, which is negligibly
small.\footnote{Superheavy topological defects can be created,
however, during
inflationary phase transitions  \cite{KL,MyBook}.}

Despite all these considerations, in the present paper   (see also
\cite{LL})
we will show  that in many theories, including the $SU(5)$
Coleman-Weinberg
theory, monopoles, as well as other topological defects,  do expand
exponentially and can be copiously produced during inflation. The
main reason
is very simple. In the center of a  monopole the scalar field $\phi$
always
vanishes. This means that monopoles always stay on the top of the
effective
potential at $\phi = 0$. When inflation begins, it makes the field
$\phi$
almost homogeneous and very close to zero near the center of the
monopole. This
provides excellent conditions for inflation inside the monopole. If
inflation
is supported by the vacuum energy $V(0)$, the conditions for
inflation  remain
satisfied inside the monopoles even after inflation finishes outside
them. An
outside observer will consider such monopoles as magnetically charged
Reissner-Nordstr\"om black holes. A typical distance between such
objects will
be exponentially large, so they will not cause any cosmological
problems.
However, each such magnetically charged black hole will have an
inflationary
universe inside it.  And, as we are going to argue, each such
inflationary
universe will contain many other inflating monopoles, which after
inflation
will look like black holes with inflationary heart. We will call such
configurations {\it fractal monopoles}, or, more generally, {\it
fractal
topological defects}.

\

2.  To explain the basic idea in a more detailed way, we will
consider first
the simple model with the Lagrangian
\begin{equation}\label{1}
L= {1\over 2} (\partial _\mu \phi )^2 - {\lambda\over 4} \Bigl(\phi^2
  - {m^2\over \lambda} \Bigr)^2  \ ,
\end{equation}
where  $\phi$ is a real scalar field.
Symmetry breaking in this model leads to formation of domains with
$\phi = \pm
\eta$, where $\eta = {m\over \sqrt \lambda}$. These domains are
divided by
domain walls   (kinks) which interpolate between
the two minima. Let us for a moment neglect gravitational effects. In
this case
one can easily obtain a solution for a static domain wall  in the
$yz$ plane:
\begin{equation}\label{2}
\phi = \eta\ {\tanh} \Bigl( \sqrt{\lambda\over 2}\ \eta x  \Bigr) \ .
\end{equation}

Now let us see whether our neglect of gravitational effects is
reasonable. The
potential energy density in the center of the kink (\ref{2}) at $x =
0$ is
equal to ${\lambda\over 4}  \eta ^4$, the gradient energy is also
equal to
${\lambda\over 4}  \eta ^4$. This energy density remains almost
constant  at
$|x| \ll m^{-1} \equiv {1\over \sqrt \lambda \eta}$, and then it
rapidly
decreases.  Gravitational effects can be neglected if the
Schwarzschild radius
$r_g = {2M\over M_P^2}$ corresponding to the distribution of matter
with energy
density $\rho = {\lambda\over 2}  \eta ^4$ and radius $R \sim m^{-1}$
is much
smaller than $R$. Here $M = {4\pi\over 3} \rho R^3$. This condition
suggests
that gravitational effects can be neglected for $\eta \, \ll \,
{3\over 2\pi}\,
M_p$. In the opposite case,
\begin{equation}\label{3}
\eta \, \ga \, {3\over 2\pi}\, M_p  \ ,
\end{equation}
gravitational effects can be very important.

This result remains true for other topological defects as well. For
example,
recently it was shown that monopoles in the theory with the scale of
spontaneous symmetry breaking $\eta \ga M_P$ form
Reissner-Nordstr\"om black
holes \cite{Lee}. It remained unnoticed, however, that the same
condition,
$\eta \ga M_P$, is simultaneously a condition of inflation  at $\phi
\ll \eta$
in the model (\ref{1}).

Indeed, inflation occurs at $\phi \ll \eta$ in the model (\ref{1}) if
the
curvature of the effective potential $V(\phi)$ at $\phi \ll \eta$ is
much
smaller than $3 H^2$, where $H = \sqrt{2\pi\lambda\over 3}\,
{\eta^2\over M_P}$
is the Hubble constant supported by the effective potential
\cite{MyBook}. This
gives
$m^2 \ll {2\pi\lambda \eta^4/ M_P^2}$, which leads to the condition
almost
exactly coinciding with (\ref{3}): $\eta \, \gg \,  M_p/\sqrt{2\pi}$.

This coincidence by itself does not mean that domain walls and
monopoles in the
theories with $\eta \, \gg \,  M_p/\sqrt{2\pi}$ will inflate. Indeed,
inflation
occurs only if the energy density is dominated by the vacuum energy.
As we have
seen, for the wall (\ref{2}) this was not the case:  gradient energy
density
for the solution (\ref{2}) near $x = 0$  is equal to the potential
energy
density. However, this is correct only after inflation and only if
gravitational effects are not taken into account.

At the initial stages of inflation the field $\phi$ is equal to zero.
Even if
originally there were any gradients of this field, they rapidly
become
exponentially small. Each time $\Delta t = H^{-1}$ new  perturbations
with the
 amplitude ${H/\sqrt{2} \pi}$ and the wavelength $\sim H^{-1}$ are
produced,
but their gradient energy density $\sim H^4$ is always much smaller
than
$V(\phi)$ for $V(\phi) \ll M^4_P$ \cite{LLM}.

Formation of domain walls in this scenario can be explained as
follows. Assume
that in the beginning the field $\phi$ inside some domain of initial
size
$O(H^{-1})$ was equal to zero. It did not move classically at that
time, since
$V'(0) = 0$.  However, within the time $\Delta t = H^{-1}$
fluctuations of the
field $\phi$ were generated, which looked like sinusoidal waves with
an
amplitude ${H/\sqrt{2} \pi}$ and wavelength $\sim H^{-1}$. During
this time the
original domain grows in size $e$ times, its volume grows $e^3 \sim
20$ times.
Therefore it becomes divided into $20$ domains of a size of the
horizon
$H^{-1}$. Evolution of the field inside each of them occurs
independently of
the processes in the other domains (no-hair theorem for de Sitter
space). In a
half of these domains scalar field will have   average value $\phi
\approx  +
{H/ {2} \pi}$, in other domains it will have   average value $\phi
\approx  -
{H/ {2} \pi}$. After that, the field $\phi$   begins its classical
motion. In
those   domains where the resulting amplitude of the fluctuations is
positive,
the field $\phi$ moves towards the minimum of the effective potential
at $\phi
= + \eta$; in other domains it moves towards $\phi = - \eta$. These
domains
become separated by the domain walls with $\phi = 0$. However, as we
already
mentioned, at this stage the gradient energy near these walls is
similar to
$H^4 \sim  V^2(\phi)/M_P^4$, which is much smaller than $V(\phi)$ for
$V(\phi)
\ll M^4_P$. Thus, domain walls continue expanding exponentially {\it
in all
directions}.

Domain wall formation is not completed at this point yet. Indeed, the
field
$\phi$ may jump back from the region with $\phi > 0$ to $\phi < 0$.
However,
the probability of such jumps is smaller than $1/2$, and it becomes
even much
smaller with a further growth of $|\phi|$. What is more important,
the typical
wavelength of new fluctuations remains $H^{-1}$, whereas the
previously formed
domains with $\phi \sim  \pm {H/ {2} \pi}$ continue growing
exponentially.
Therefore the domains with negative $\phi$ produced by the jumps of
the field
$\phi$ will appear inside   domains with positive $\phi$, but they
will look
like   small islands with $\phi < 0$ inside the sea with $\phi >0$.
This means
that the subsequent stages of inflation cannot destroy the originally
produced
domain walls; they can only produce new domain walls on a smaller
length scale.
These new walls will be formed only in those places where the scalar
field is
sufficiently small for the jumps with the change of the sign of the
field
$\phi$ to be possible. Therefore the new walls will be created
predominantly
near the old ones (where $\phi = 0$), thus forming a fractal domain
wall
structure.

\

3. Similar effects occur in more complicated models where instead of
a
discrete symmetry $\phi \to - \phi$ we have a continuous symmetry.
For example,
instead of the model (\ref{1}) one can consider a model
\begin{equation}\label{5}
L= \partial _\mu \phi^*\,  \partial _\mu \phi  - {\lambda}
\left(\phi^*\phi
  - {\eta^2\over 2 } \right)^2  \ ,
\end{equation}
where  $\phi$ is a complex scalar field, $\phi = {1\over \sqrt
2}(\phi_1 + i
\phi_2)$.
Spontaneous  breaking of the $U(1)$ symmetry in this theory may
produce global
cosmic strings. Each string contains  a line with $\phi = 0$. Outside
this line
the absolute value of the field $\phi$ increases and asymptotically
approaches
the limiting value  $\sqrt{\phi_1^2 + \phi_2^2} = \eta$. This string
will be
topologically stable if the isotopic vector $(\phi_1(x), \phi_2(x))$
rotates by
$2n\pi$ when the point $x$ takes a closed path  around the string.

The next step is to consider a theory with $O(3)$ symmetry,
\begin{equation}\label{6}
L= {1\over 2}  (\partial _\mu\vec \phi )^2 -
{\lambda \over 4} ( {\vec \phi}^2 - {{\eta ^ 2}} ) ^2
  \ ,
\end{equation}
where $\vec \phi$ is a vector $(\phi_1, \phi_2, \phi_3)$. This theory
admits
global monopole solutions. The simplest monopole configuration
contains a point
$x = 0$ with $\phi(0) = 0$ surrounded by the scalar field $\vec
\phi(x) \propto
\vec x$. Asymptotically this field approaches regime with $\vec
\phi^2(x) =
\eta^2$.

The most important feature of strings and monopoles is the existence
of the
points where $\phi = 0$. Effective potential has an extremum at $\phi
= 0$, and
if the curvature of the effective potential is smaller than $H^2 =
{8\pi
V(0)\over 3M^2_P}$, space around the points with $\phi = 0$ will
expand
exponentially, just as in the domain wall case considered above.

Now we will add gauge fields. We begin with the Higgs model, which is
a direct
generalization of the model (\ref{5}):
\begin{equation}\label{5a}
L= D _\mu \phi^*\,  D _\mu \phi  -  {1\over 4} F _{\mu \nu} F^{\mu
\nu}   -
{\lambda} \left(\phi^*\phi
  - {\eta^2 \over 2 }\right)^2  \ .
\end{equation}
Here $D _\mu$ is a covariant derivative of the scalar field, which in
this
simple case (Abelian theory) is given by $\partial _\mu - i e A_\mu$.
In this
model strings of the scalar field contain  magnetic flux $\Phi =
2\pi/e$. This
flux is localized near the center of the string with $\phi(x) = 0$,
for the
reason that the vector field becomes heavy at  large $\phi$, see e.g.
\cite{Kirzhnits}. However, if inflation  takes place inside the
string, then
the field $\phi$ becomes vanishingly small not only at the central
line with
$\phi(x) = 0$, but even exponentially far away from it. In such a
situation the
flux of magnetic field will not be confined near the center of the
string. The
thickness of the flux will grow together with the growth of the
universe. Since
the total flux of magnetic field inside the string is conserved, its
strength
will decrease exponentially,
and very soon its effect on the  string expansion will become
negligibly small.
A deep underlying reason for this behavior is the conformal
invariance of
massless vector fields. Energy density of such fields decreases as
$a(t)^{-4}$,
where $a(t)$ is a scale factor of the universe. The final conclusion
is that
the vector fields do not prevent inflation of strings. The condition
for
inflation to occur inside the string remains the same as before: the
curvature
of the effective potential at $\phi = 0$ should be smaller than $H^2
= {8\pi
V(0)\over 3M^2_P}$.

The final step is to consider  magnetic monopoles. A simplest example
is given
by the $O(3)$ theory
\begin{equation}\label{7}
L= {1\over 2}  |(D _\mu {\vec \phi} | ^2  - {1\over 4} F^a _{\mu \nu}
F^{a \mu
\nu}-
{\lambda \over 4} ( {\vec \phi}^2 - {{\eta ^ 2}}  ) ^2  \ .
\end{equation}
Global monopoles of the theory (\ref{5}) become magnetic monopoles in
the
theory (\ref{5a}). They also have $\phi = 0$ in the center.  Vector
fields in
the center of the monopole are massless ($g\phi = 0$). During
inflation these
fields exponentially decrease, and therefore they do not affect
inflation of
the monopoles.

We should emphasize that even though the field $\phi$ around the
monopole
during inflation is very small, its topological charge is well
defined,  it
cannot change and it cannot annihilate with the charge of other
monopoles as
soon as the radius of the monopole becomes greater than $H^{-1}$.
However, an
opposite process is possible. Just as domain walls can be easily
produced by
quantum fluctuations near other inflating domain walls, pairs of
monopoles can
be produced in the vicinity of an inflationary monopole. The distance
between
these monopoles grow exponentially, but the new monopoles will appear
in the
vicinity of each of them. This picture has been confirmed by the
results of
computer simulation of this process \cite{LL}.

Note that in the simple models discussed above inflation of monopoles
occurs
only if spontaneous symmetry breaking is extremely strong, $\eta \ga
M_P$.
However, this is not a necessary condition. Our arguments remain
valid for {\it
all} models where the curvature of the effective potential near $\phi
= 0$ is
smaller than the Hubble constant supported by $V(0)$. This condition
is
satisfied by all models which were originally proposed for the
realization of
the new inflationary universe scenario.
In particular, the monopoles in the $SU(5)$ Coleman-Weinberg theory
also should
expand exponentially. However, to make sure that we do not miss
something
important, we should resolve the paradox     formulated in the very
beginning
of this paper: The Hubble constant $H$ during inflation in the
$SU(5)$
Coleman-Weinberg theory is much smaller than the mass of the vector
field
$M_X$, which is usually related to the size of the monopole. In such
a
situation inflation of the interior of the monopole does not seem
possible.

This objection is very similar to the argument that the domain wall
should not
inflate since its gradient energy density is equal to its potential
energy
density. The resolution of the paradox is very similar. The effective
mass of
the vector field $M_X  \sim g\eta \sim 10^{15}$ GeV can determine  an
effective
size of the monopole only {\it after} inflation.   Effective mass of
the vector
field $M_X(\phi) \sim g\phi$ is always equal to zero in the center of
the
monopole. Once inflation begins in a domain of a size $O(H^{-1})$
around the
center of the monopole, it expels vectors fields away from the center
and does
not allow them to penetrate back as far as inflation continues.

\

4. It is well known that in many versions of inflationary theory some
parts of
the universe expand without end \cite{b51}--\cite{b19}. Our results
reveal a
very nontrivial role which topological defects may play in this
process.  For
example, in the theory with breaking of a discrete symmetry $\phi \to
- \phi$
the universe in the source of its evolution   becomes divided into
many
thermalized regions divided by exponentially expanding domain walls
unceasingly
producing new inflating walls.
It is important that inflation {\it never stops} near the walls. Some
of these
walls could collapse, eating the island of thermalized phase inside
them.
However, it does not seem  possible. Due to the no-hair theorem for
de Sitter
space, the part of the universe near the wall (i.e. near $\phi = 0$)
lives by
its own laws and continue expanding exponentially all the time.
Thus, the
walls become indestructible  sources of eternal inflation, which
produce new
inflating walls, which produce new walls, etc. Far away from the
walls, the
field $\phi$ relaxes near $\phi = \pm \eta$.  However, due to  the
exponential
expansion of space near the walls, they always remain exponentially
thick and
never approach the thin wall solution described by eq. (\ref{2}).

The picture of eternally inflating universe consisting of  islands of
thermalized phase surrounded by inflating domain walls is very
similar to a
picture which appears in old inflation and in the versions of new
inflation
where the state $\phi = 0$ is sufficiently stable
\cite{b51}--\cite{b62}.
However,  in our model the main reason  for this behavior is purely
topological \cite{LL}. If inflation in this theory begins at large
$\phi$, as
in the simplest models of chaotic inflation, then the universe will
not consist
of islands of thermalized phase surrounded by de Sitter space. On the
contrary,
it will consist of islands of inflationary universe surrounded by the
thermalized phase \cite{LLM}. Nevertheless, inflation in this case
will go
eternally due to the process of self-reproduction of inflationary
domains with
large $\phi$ \cite{b19}.

 The results of computer simulations performed in \cite{LL} suggest
that in the
theories with continuous symmetry breaking, which allow existence of
inflating
strings and monopoles, the global structure of the universe is
similar to the
structure of the universe in the simplest versions of chaotic
inflation. The
universe will consist of islands of inflationary phase associated
with
inflating monopoles (or of lines of inflationary phase associated
with
inflating strings) surrounded by thermalized phase. Since   monopoles
are
topologically stable, and field $\phi$ is equal to zero in their
centers,  they
form indestructible sources of inflation. The structure of space-time
near each
such monopole is very complicated; it should be studied by the
methods
developed in \cite{Berezin} for description of a bubble of de Sitter
space
immersed into vacuum with vanishing energy density.  Depending on
initial
conditions, many possible configurations may appear. The simplest one
is the
monopole which looks like a small magnetically charged black hole
from the
outside, but which contains a part of exponentially expanding space
inside it.
This is a wormhole configuration similar to those   studied in
\cite{Markov}--\cite{Tkachev}. The physical interpretation of this
configuration can be given as follows \cite{Berezin,Tkachev}. An
external
observer will   see a small magnetically charged Reissner-Nordstr\"om
black
hole.  This part of the picture is consistent with the results
obtained in
\cite{Lee}. On the other hand,   an observer near  the monopole  will
see
himself inside an eternally inflating part of the universe. These two
parts of
the universe will be connected by a wormhole. Note that the wormhole
connecting
  inflationary magnetic monopole to our space cannot evaporate unless
the
inflationary universe can loose its magnetic charge.

At the quantum level the situation becomes even more interesting. As
we already
mentioned, fluctuations of the field $\phi$ near the center of a
monopole are
strong enough to create new regions of space with $\phi = 0$, some of
which
will become   monopoles. After a while, the distance between these
monopoles
becomes exponentially large, so that they cannot annihilate. This
process of
monopole-antimonopole pair creation produces a fractal structure
consisting of
monopoles created in the vicinity of other monopoles.  Each of these
monopoles
in the process of its further evolution will evolve into   a black
hole
containing inflationary universe containing many other monopoles,
etc. That is
why we call them {\it fractal monopoles}.

Note that one of the main motivations for the development of
inflationary
cosmology was a desire to get rid of primordial monopoles. However,
the
original version of the new inflationary universe scenario based on
the theory
of high temperature phase transitions did not work particularly well,
since it
required rather unnatural initial conditions \cite{MyBook}.
Fortunately, the
same potentials which gave rise to new inflation can be used in the
context of
the chaotic inflation scenario, which does not require the whole
universe being
in a state of thermal equilibrium with $\phi = 0$. If these initial
conditions
can be realized at least in one domain of initial size $O(H^{-1})$,
the
universe enters the stage of eternal inflation. Of course, this
possibility is
still rather problematic since in the models used in the new
inflationary
universe scenario the size $O(H^{-1})$ is  many orders of magnitude
greater
than the Planck length. For example, the probability of  quantum
creation of
inflationary universe in the new inflationary universe scenario is
suppressed
by an infinitesimally small factor $\exp\Bigl(-{3M^4_P\over
8V(0)}\Bigr) \sim
10^{-10^{16}}$ \cite{Creation}. Therefore it is much easier for
inflation to
begin and continue eternally in the simplest chaotic inflation models
where
inflation is possible even at $H \sim M_P$, $V(\phi) \sim M^4_P$
\cite{LLM,b19}. Still the problem of initial conditions in the models
with the
potentials used in the new inflationary scenario does not look that
bad if one
takes into account that inflation in these models also goes on
without end
\cite{b52,b62}, and eventually we  become generously rewarded for our
initial
problems by the infinite growth of  the total volume of  inflationary
domains.
Moreover, as it was shown in \cite{LLM}, the problem  of initial
conditions for
the models with the potentials used in the new inflationary scenario
can be
easily solved by a preceding stage of chaotic inflation. Now we
encounter a new
amazing twist of   this scenario.  Eternal inflation begins if
initially we
have at least one inflating monopole. The possibility that monopoles
by
themselves solve the monopole problem is so simple that it certainly
deserves
further investigation.

It is a pleasure to thank  Victor Berezin, Valery Frolov, Renata
Kallosh and
Igor Tkachev  for valuable discussions.  Many ideas discussed in this
paper
have been provoked by the results of computer simulations of
stochastic
properties of inflationary universe performed together with Dmitri
Linde
\cite{LL}. After this work has been completed, I was informed by Alex
Vilenkin
that he  recently obtained similar results. This work was supported
in part  by
NSF grant PHY-8612280.

\end{document}